# Immobilization of single strand DNA on solid substrate

S. A. Hussain[$^], P. K. Paul[$], D. Dey[$], D. Bhattacharjee[$] and S. Sinha[&]

[$]Department of Physics, Tripura University, Tripura, INDIA
[&]Department of Life Science, Tripura University, Tripura, INDIA

**Abstract:**
Thin films based on Layer-by-Layer (LbL) self assembled technique are useful for immobilization of DNA onto solid support. This communication reports the immobilization of DNA onto a solid support by electrostatic interaction with a polycation Poly (allylamine hydrochloride) (PAH). UV-Vis absorption and steady state fluorescence spectroscopic studies exhibit the characteristics of DNA organized in LbL films. The most significant observation is that single strand DNA are immobilized on the PAH backbone of LbL films when the films are fabricated above the melting temperature of DNA. DNA immobilized in this way on LbL films remains as such when the temperature is restored at room temperature and the organization remains unaffected even after several days. UV-Vis absorption spectroscopic studies confirm this finding.

**Keywords:**
Layer-by-Layer (LbL) self assembled films, UV-Vis absorption spectroscopy, DNA melting, Immobilization of DNA.

^ **e-mail:** sa_h153@hotmail.com

## 1. Introduction:

DNA (deoxyribonucleic acid) is an essential biological material whose base sequence controls the heredity of life. DNA is also an interesting anionic polyelectrolyte with unique double helix structure [1]. On the basis of hydrogen bonding properties of the DNA base pairs, oligonucleotide probes were designed in recent time to detect tumor gene and various biosensors were also proposed [2, 3]. Immobilization of DNA onto a solid substrate has profound biological and technological advantages starting from DNA chips to single molecule biology [4]. It is also important in a wide range of research areas including work on nanoparticles [5], DNA computing [6, 7], and DNA chip technologies. Some of the common substrates for the immobilization of DNA are latex particles [8], glass surface [9], silanized mica surface [9], gold nanoparticles [10], polystyrene microspheres [8] etc. The requirement for reproducible stable surfaces has created increasing interest in designing modified surfaces to immobilize DNA.

The techniques used for immobilization of DNA onto solid substrates are Layer-by-Layer (LbL) self assembled by electrostatic interaction [11] as well as also by Langmuir-Blodgett (LB) technique [12].

LbL films are predominantly assembled by the alternate deposition of positively and negatively charged polymers with film build up facilitated by electrostatic interaction [13, 14].

In one of the previous studies on DNA-LbL films [15], DNA strands were held together by hydrogen bonding of the base pairs, and the control of the morphology of the film was achieved by manipulating the extent of hydrogen bonding and repulsion between the negatively charged phosphate backbones. In another work hydrogen bonding and hydrophobic interaction were used to facilitate LbL assembly of uncharged polymers [11].

Here we have studied the photophysical characteristics of DNA LbL films fabricated by electrostatic interaction of polyanionic DNA with a polycationic poly (allylamine hydrochloride) (PAH). PAH being a photophysically inert substance, UV-Vis absorption and steady state fluorescence spectroscopic studies exhibit the characteristics of DNA organized in LBL films. The films fabricated by changing various parameters have been subjected to our investigation.

The thermal separation of DNA strands which is commonly known as the denaturation or melting [1] has been experimentally observed by taking the UV-Vis absorption spectra of aqueous solution of DNA at different temperatures. The specificity of hydrogen bonding between the bases along with the phosphodiester bond maintain the basic double helical structure of DNA and at higher temperature due to the increase in thermal energy, breaking of only hydrogen bonds occur. As a result the two DNA strands are separated. The denaturation of DNA changes the intensity of its absorption band at 260 nm. Intensity of 260 nm band in single stranded DNA solution is much higher than that of the double stranded DNA solution [1].

## 2. Experimental:

Herring sperm sheared DNA purchased from SRL India, was used as received. The purity of DNA was checked by UV-Vis absorption and fluorescence spectroscopy before use. Poly (allylamine hydrochloride) (PAH) was used as the polycation. PAH (molecular weight = 70,000), purity > 99%, were purchased from Aldrich Chemical Co. and were used without any further purification. The deposition bath was prepared with 0.2 mg/ml DNA and $10^{-3}$ M (based on the repeat units for polyion) of PAH aqueous solutions using triple distilled deionised (18.2 MΩ) Millipore water.

Layer-by-layer (LbL) self-assembled films were obtained by dipping thoroughly cleaned fluorescence grade quartz substrates alternately in solutions of the polyelectrolyte PAH and oppositely charged DNA. PAH was used as polycation for the fixation of the DNA molecule to the substrate. First of all the quartz substrate was cleaned by standard procedure [13] and immersed in the PAH solution for 15 minutes followed by rinsing in water bath for 2 minutes. The rinsing washes off the surplus cation attached to the surface. The substrate was then immersed in DNA solution for 15 minutes followed by same rinsing procedure. After each deposition and rising procedure sufficient time was allowed to dry up the film and their UV-Vis absorption spectra were recorded to monitor the film growth. Deposition of the PAH (cation) and DNA (anion) layers resulted in one bilayer of self-assembled PAH-DNA LbL film. The whole sequence of the film deposition procedure was repeated for the preparation of desired number of bilayer LbL films.

Denaturation of DNA was studied by dipping one layer PAH film into DNA solution at different temperature, keeping, the immersion time in DNA solution constant for 15 minute.

The UV-Vis absorption and steady state fluorescence spectra of the LBL films as well as solutions were recorded using Lambda-25 UV-Vis spectrophotometer, Perkin Elmer and LS-55 fluorescence spectrophotometer, Perkin Elmer respectively.

## 3. Results and discussion:

Figure 1 shows the UV-Vis absorption and steady state fluorescence spectra of the aqueous solution of DNA ($10^{-3}$ M), aqueous solution of DNA-PAH mixture (1:1 volume ratio) and DNA microcrystal.

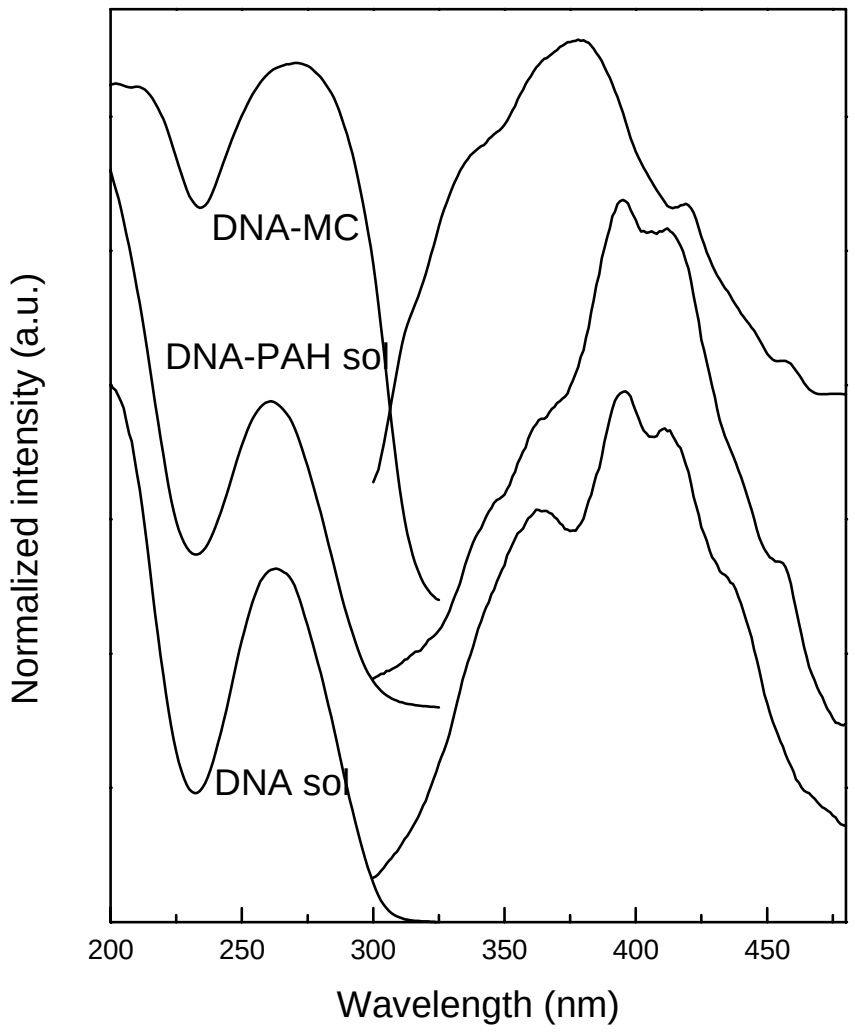

***Figure 1:*** *UV-Vis absorption and steady state fluorescence spectra of pure DNA, DNA-PAH (1:1 volume ratio) solution and pure DNA microcrystal*

The absorption spectrum of aqueous solution of DNA shows a distinct and broad band with an intense peak at 262 nm which is the characteristics of DNA absorption and is consistent with the reported result [11]. This band is due to the absorption of light by the nucleic acid bases [1] of DNA and the transition dipole moment is directed along the long axis of DNA strands.

Absorption spectrum of DNA-PAH mixed aqueous solution shows similar band pattern. This is due to the fact that DNA interacts with the cationic $NH_3^+$ group of PAH molecule through its phosphate ($PO_4^-$) ion tagged at the sugar base, whereas the purine and pyrimidine bases remain unaffected.

DNA microcrystal absorption spectrum gives broadened absorption profile with almost similar band pattern. However the peak position is shifted to about 270 nm. This may be due to the change of microenvironment of individual DNA molecules in the microcrystal and in solution or in films. It is interesting to mention in this context is that A. Bhaumik et. al. [ref] reported that several DNA strands overlap together to form bundles on the surface of the Langmuir-Blodgett films.

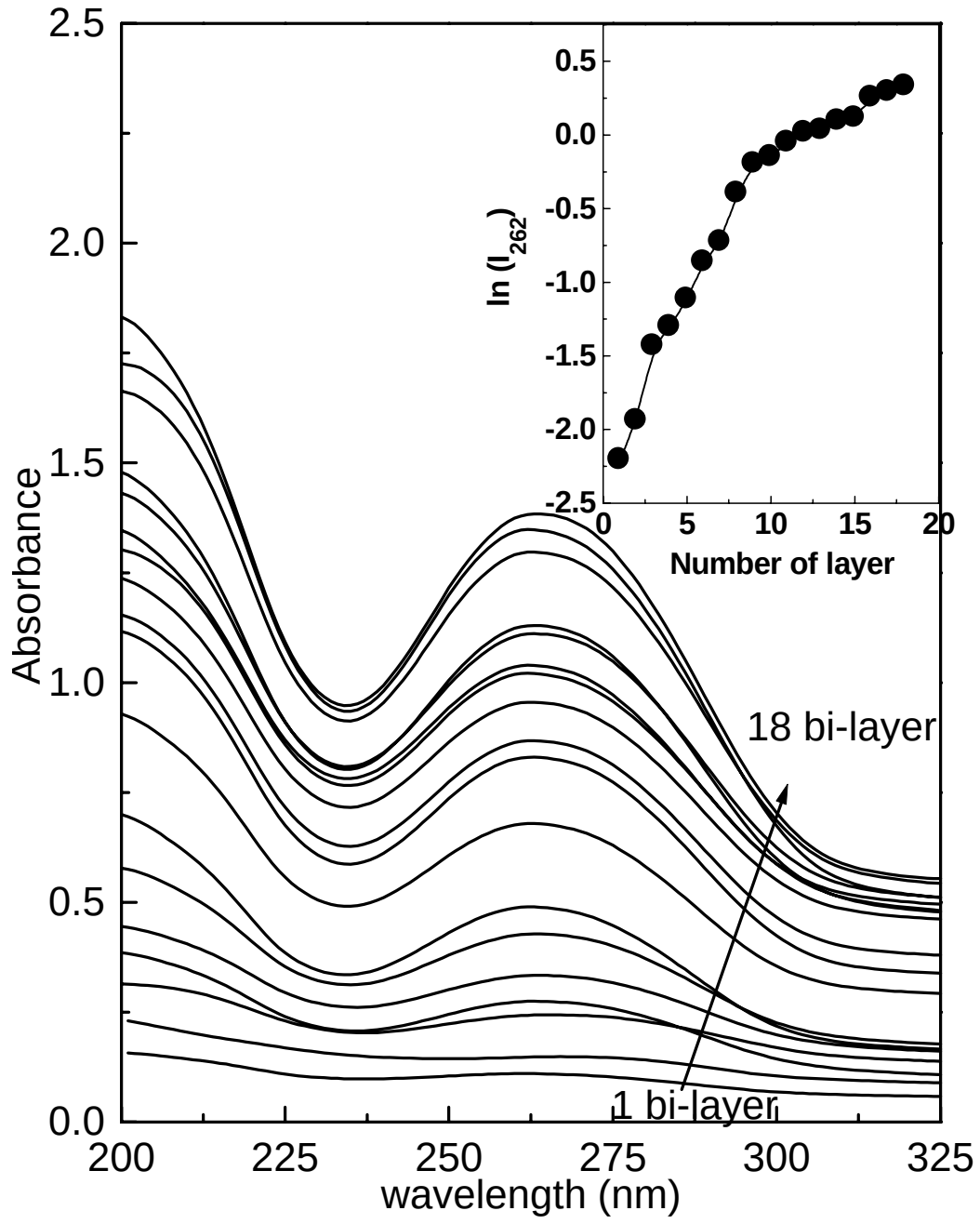

***Figure 2a:*** *UV-Vis absorption spectra of different layered (1-33 bilayer) DNA-PAH Layer-by-Layer self assembled films. Inset shows the plot of intensity of 262 nm band as a function of layer number.*

Fluorescence spectrum of aqueous solution of DNA shows prominent vibrational bands with intense peak at 396 and 412 nm along with a weak peak at high energy region of 362 nm. This structure is almost similar in case of DNA-PAH solution. However, in the DNA microcrystal spectrum a broad band is observed along with a diffused vibrational structure.

Figure 2a shows the absorption spectra of different layered PAH-DNA LbL films starting from 1 bi-layer to 33 bi-layer. In all the cases absorption bands give the similar spectral profile with peak at 262 nm. This is a clear indication that DNA molecules are successfully incorporated into the PAH-DNA LbL films due to the interaction with PAH molecules.

The inset of figure 2a shows the plot of integrated intensity of absorption maximum (262 nm) as a function of layer number. From the figure it was observed that the intensity of this band increases linearly for lower number of layers (up to 10 layer) indicating uniform deposition of DNA molecule onto quartz substrate. However for higher number of layer (> 10 layer) the intensity does not increases linearly. This may be due to the fact that at higher number of layers the DNA deposition onto quartz substrate is not uniform and certain amount of material loss is occurred. That is few DNA molecules may come out off the film during PAH deposition or washing the film with pure water.

The absorption spectra of PAH-DNA one bi-layer LbL films with different dipping time is shown in figure 3. Here in all the cases the deposition time of polycation was kept fixed at 15 minutes. However the DNA deposition times were taken from one minute and by varying different time interval up to a maximum time limit of fifty minutes. From the figure it has been observed that the intensity of absorption spectra increases initially with DNA deposition time and gets saturation for DNA deposition time of twelve minutes. This is evident from the plot of the intensity of absorption maximum versus time (inset of figure 3) of 262 nm. peak. This is a clear indication that the interaction of DNA molecules with the PAH layer was completed by twelve minutes and no PAH molecule remains free after twelve minutes within the film for further interaction with the DNA molecules.

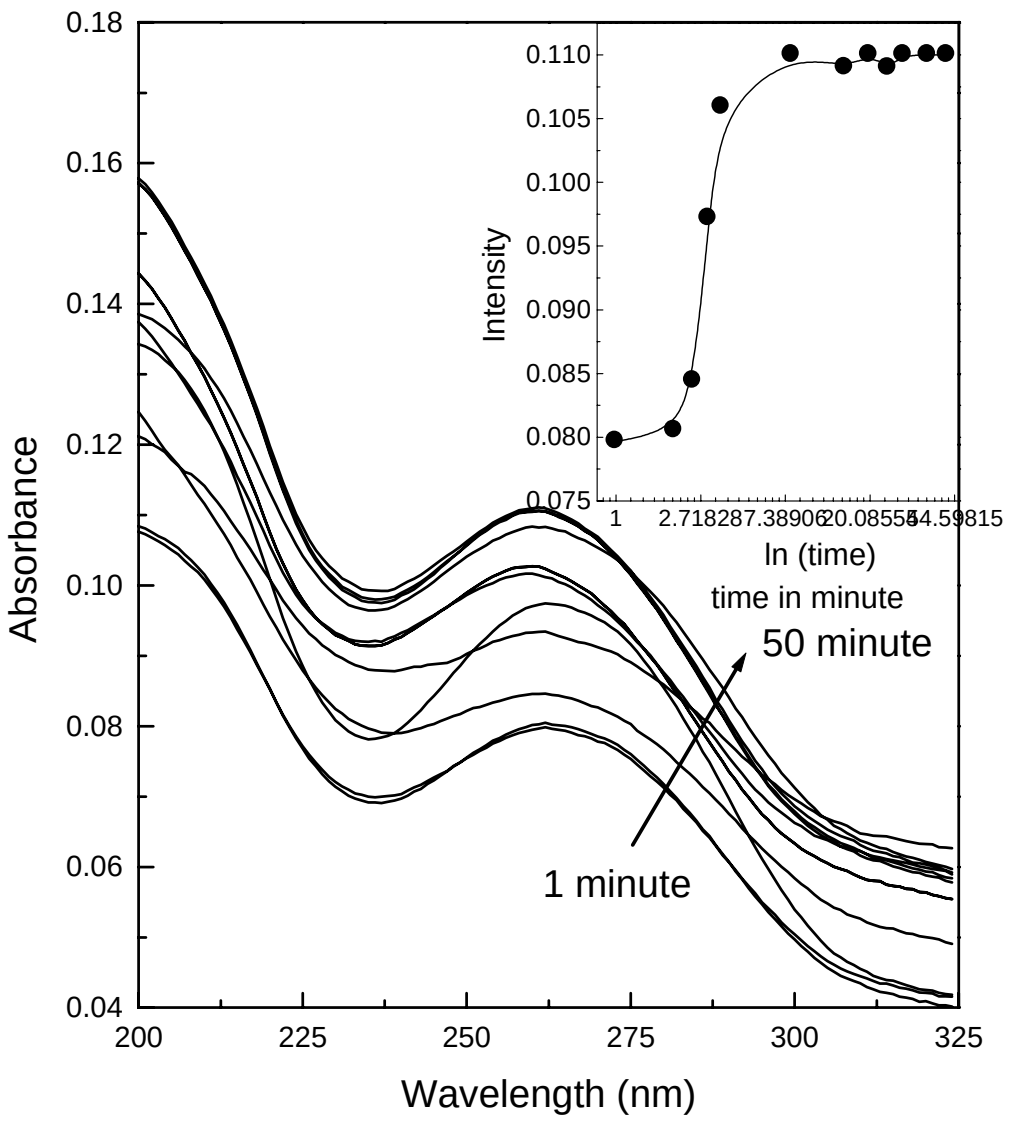

***Figure 3:*** *UV-Vis absorption spectra of 1-bilared DNA-PAH Layer-by-Layer self assembled film with PAH deposition time 15 minute and different DNA deposition times (1-50 minutes). Inset shows the plot of intensities of 262 nm band as a function of time (logarithmic scale).*

It is a well-known fact that at and above $60^0$C, breaking of hydrogen bonds of the double stranded DNA starts and around $90^0$C DNA denaturation or melting [1] occurs resulting in complete separation of two strands.

In order to check whether the two DNA strands can be immobilized, we performed a unique experiment of fabricating the LbL films of DNA at various temperatures on a preassembled PAH monolayer onto quartz substrate.

DNA monolayer was deposited on PAH monolayer at different temperature starting from $60^0$C up to $95^0$C with an interval of $5^0$C. The dipping time was kept fixed at 15 minute.

UV-Vis absorption spectra of these LbL films are shown in figure 4. The most interesting part of this observation is that although the band pattern are similar in all the cases, the absorbance intensity increases to a large extent starting from the films fabricated at $60^0$C to the film fabricated at $90^0$C. Around $90^0$C temperature the absorbance intensity becomes maximum in our experimental design.

With further increase in temperature to $95^0$C, the absorbance intensity is abruptly reduced. Films cannot be fabricated above $95^0$ C temperature due to technical difficulties. Films fabricated by reversing the temperature cycle also give the similar absorbance intensity pattern as shown in figure 4 (dotted line).

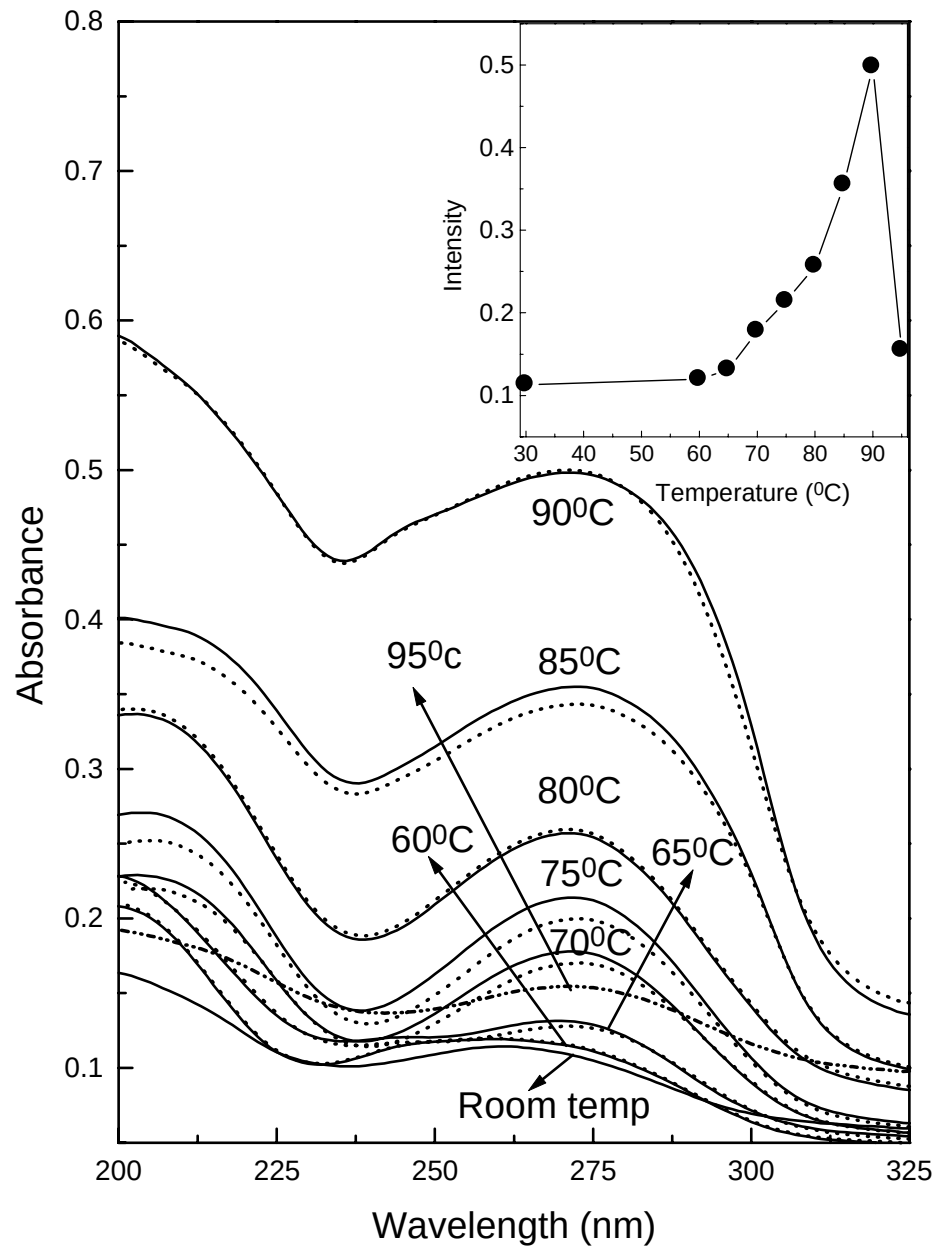

***Figure 4:*** *UV-Vis absorption spectra of 1-bilayered DNA-PAH layer-by-layer self assembled films fabricated at various temperature viz. $60^0C$, $65^0C$, $70^0C$, $75^0C$, $80^0C$, $85^0C$, $90^0C$ and $95^0C$ on a preassembled PAH layer. Solid lines represent fabrication at increasing temperature; dotted lines represent fabrication at decreasing temperature. Inset shows the plot of intensity of 262 nm band as a function of time.*

Figure 5a shows the arrangement of DNA molecules in the PAH-DNA LbL films formed at room temperature where the initial double helix structure of DNA prevails and the two strands of DNA twist around each other forming the so called double helix. As a result the length of the individual strand of the DNA helix gets reduced. In the light of this knowledge, the absorbance intensity of the DNA is reduced at a lower temperature, since the 262 nm absorption band originates along the long axis transition. Above $60^0$C temperature the hydrogen bonds started breaking due to thermal agitation and the two DNA strands started to be separated away from each other. In the films fabricated at that higher temperature, organization of DNA molecules in the PAH-DNA LbL films is shown in figure 5b, the axial length of the DNA strand increases consequently absorbance intensity is also increased.

However after melting which is around $90^0$C temperature, the two strands of DNA become totally separated. The single stranded DNA molecules resulting from denaturation, form random coil without a regular structure. In the films fabricated above denaturation temperature, the organizations of DNA molecules on the PAH backbone of the solid substrate are shown in figure 5c, axial length of DNA strands is sufficiently reduced and as a consequence the absorbance intensity abruptly falls.

DNA molecules once immobilized at higher temperature in the PAH-DNA mixed LbL films, their organization remain unaffected even if the films are taken back to normal temperature, as is evidenced from the absorbance intensity shown in figure 4.

Stabilization studies also indicate that once the DNA are immobilized on the PAH films in different organization, they remain so even after fifteen days (figure not shown).

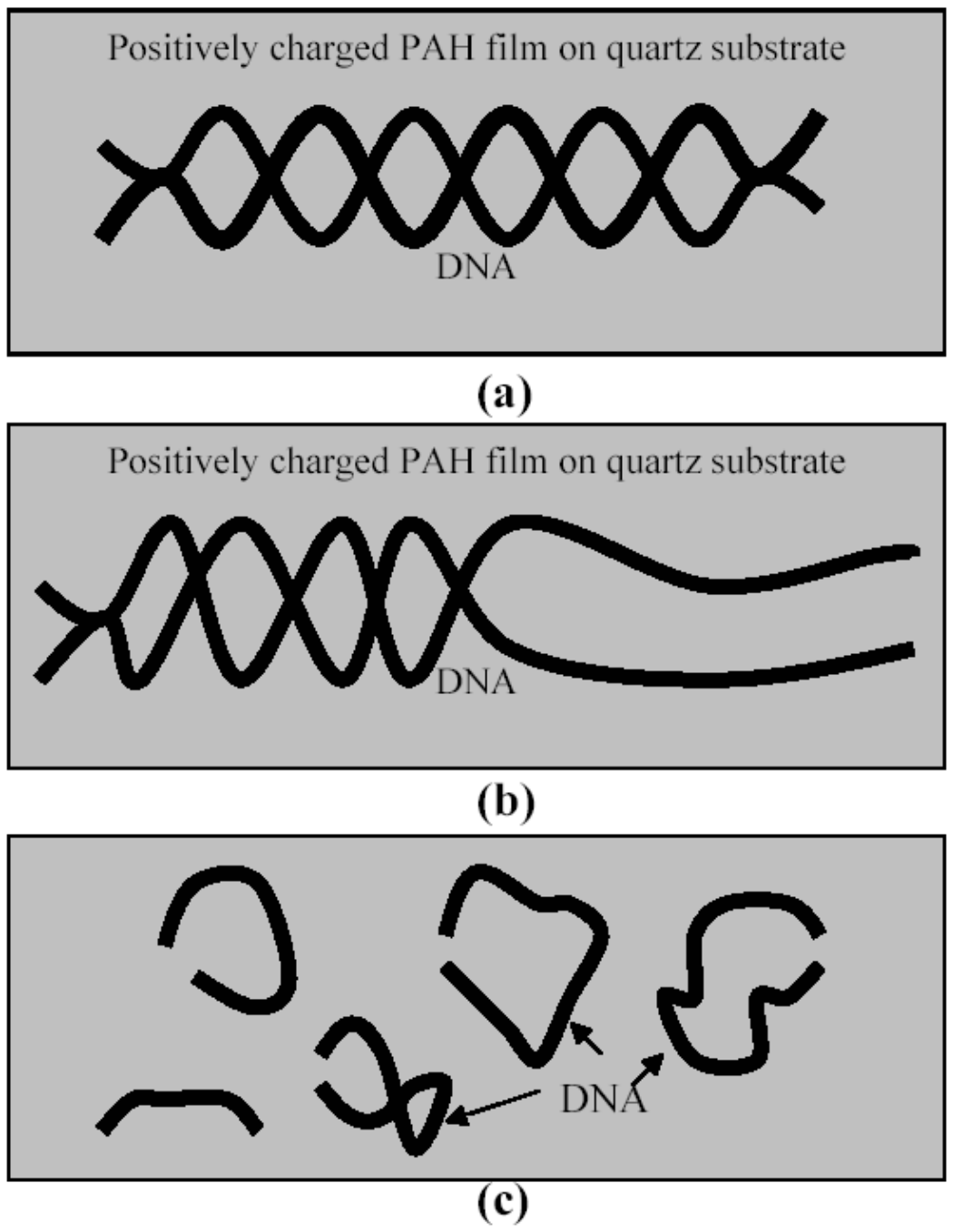

***Figure 5:*** *Schematic representation of double strand DNA structure and their organization on LbL films at different temperature.*

The most interesting observation in our work is that once the denatured single strands are immobilized on the PAH backbone of the LbL films at higher temperature, they remain immobilized on the films even when the films are again taken back to the room temperature. This is manifested by the increase in the intensity of the absorption band in LbL films fabricated at higher temperature than that fabricated at lower temperature, although all the absorption spectra are taken at room temperature. It has also been observed that once the denatured single strand DNA were immobilized by PAH on the LbL films, the organization of DNA remains unaffected even after fifteen days as was evidenced by the unaltered absorbance intensity.

## 4. Conclusion:

In conclusion our results show that DNA molecules can be immobilized on the PAH backbone of a solid substrate by electrostatic interaction. Absorption spectroscopic studies definitely confirm the successful incorporation of DNA molecules in DNA-PAH LbL films. Spectroscopic characteristics also exhibit the nature of DNA organization in LbL films. Time dependent studies indicate that the whole process of formation of DNA layer on PAH backbone takes only 12 minute and after that time no PAH molecules remains free to interact with the excess DNA molecules within the solution. The most significant observation is that when the films are fabricated above the melting temperature of DNA, single strand DNA are immobilized on the backbone of PAH films. DNA immobilized in this way on LbL films remains so even when the temperature is restored at room temperature and the organization remains unaffected even after several days.

**Acknowledgement:**

The authors are grateful to DST and CSIR, Govt. of India for providing financial assistance through FIST-DST Project No. SR/FST/PSI-038/2002 and CSIR project Ref. No. 03(1080)/06/EMR-II